\documentclass[runningheads]{llncs}
\usepackage{graphicx}
\usepackage{hyperref}
\newcommand{\squeezeup}{\vspace{-7.0mm}}

\let\svthefootnote\thefootnote
\newcommand\freefootnote[1]{%
  \let\thefootnote\relax%
  \footnotetext{#1}%
  \let\thefootnote\svthefootnote%
}

\begin{document}
\title{Effects of interactivity and presentation on review-based explanations for recommendations}
\titlerunning{Effects of interactivity and presentation on review-based expl. for recomm.}
%

\author{Diana C. Hernandez-Bocanegra\and
Jürgen Ziegler}

\authorrunning{Hernandez-Bocanegra and Ziegler}
\institute{University of Duisburg-Essen, Duisburg 47057, Germany
\email{\{diana.hernandez-bocanegra,juergen.ziegler\}@uni-due.de}}
%
%


\maketitle

\begin{abstract}
User reviews have become an important source for recommending and explaining products or services. Particularly, providing explanations based on user reviews may improve users’ perception of a recommender system (RS). However, little is known about how review-based explanations can be effectively and efficiently presented to users of RS. We investigate the potential of interactive explanations in review-based RS in the domain of hotels, and propose an explanation scheme inspired by dialog models and formal argument structures. Additionally, we also address the combined effect of interactivity and different presentation styles (i.e. using only text, a bar chart or a table), as well as the influence that different user characteristics might have on users' perception of the system and its explanations. To such effect, we implemented a review-based RS using a matrix factorization explanatory method, and conducted a user study. Our results show that providing more interactive explanations in review-based RS has a significant positive influence on the perception of explanation quality, effectiveness and trust in the system by users, and that user characteristics such as rational decision-making style and social awareness also have a significant influence on this perception.
\end{abstract}


\section{Introduction}

Explaining \freefootnote{The final publication is available at Springer via \url{http://dx.doi.org/10.1007/978-3-030-85616-8_35}}  the recommendations generated algorithmically by a recommender system (RS) has been shown to offer significant benefits for users with respect to factors such as transparency, decision support, or trust in the system \cite{Tintarev07,Tintarev12}. Many approaches to explaining the products or services suggested by an RS have been based on ratings provided by other users or properties of the recommended items, approaches related to collaborative and content-based filtering methods \cite{Herlocker00,Vig09}. More recently, fueled by the advances in natural language processing, user-written reviews have received considerable attention as rich sources of information about an item’s benefits and disadvantages, which can be utilized for explanatory purposes. Reviews are, however, subjective, and may be inconsistent with the overall rating given by the user. Even when overcoming the challenge of processing noisy review texts, the question of which review-based information to show and how to present it is still largely open, partly due to the lack of empirical conclusions on how to best present review-based explanations, just as there is a general lack of user-centric evaluations of explanatory RS \cite{Nunes17}. 

While as yet no overall theoretical model of explainable recommendations has been established, we propose to analyze explanations through the lens of argumentation theory which has produced a wide range of models of argumentation \cite{Bentahar10}. One class of  argumentation models defines - with many variations - logical structures of argumentative elements such as claims, evidence or facts supporting a claim, rebuttals and other components. A recommendation issued by a RS can be considered a specific form of a claim, namely that the user will find the recommended item useful or pleasing \cite{Donkers20b}. The role of an explanation is thus to provide supportive evidence (or rebuttals) for this claim. Claims are, however, also present in the individual user’s rating and
opinions, which may require explaining their grounds as well, thus creating a complex multi-level argumentative structure in an explainable RS, a concern also raised in \cite{Friedrich11}. A different branch of argumentation theories \cite{Walton95} have abandoned the idea of static argumentation models and propose a dialectical approach to argumentation, focusing more on the process of exchanging arguments as part of a dialogue between two parties. This approach has been applied in the formulation of models \cite{Hilton90,Walton11,Madumal20} that take into account the social aspect of the explanatory process (an explainer transfers knowledge to an explainee) \cite{Miller18}: a set of interactions, an exchange of questions and answers. However, the practical application of these conceptual frameworks and their actual benefit from the users' perspective is yet to be determined, mostly due to the lack of user-centered evaluations of implementations based on such frameworks.

Thus, we propose an interactive approach to argumentative explanations based on reviews, which allows users to actively explore explanatory information, while providing answers to some of their potential questions at different levels of detail (i.e. why is this [item] recommended?, how customers rated [additional features]?, what was reported on [feature]?). Moreover, we provide empirical evidence of the effect that an implementation of this approach may have on users' perception, particularly in the hotels domain. More specifically, we aimed to answer:
\textbf{RQ1}: How do users perceive review-based explanations with different degrees of \textit{interactivity}, in terms of explanation quality, and of the transparency, efficiency and trust in the system?

Additionally, we also aimed to test the combined effect of explanation interactivity and different presentation styles, particularly: using only text, using a bar chart or using a table, to show, among others, the distribution of positive and negative comments on the quality of an item. Thus: 
\textbf{RQ2}: How do different \textit{presentation styles} influence users’ perception of review-based explanations with different degrees of interactivity?

Furthermore, we also addressed the influence that different user characteristics might have on the perception of the proposed approach. Regardless of its type, an explanation may not satisfy all possible explainees \cite{Sokol20} . Moreover, individual user characteristics can lead to different perceptions of a RS \cite{Knijnenburg12,Xiao07}, for which we assumed that this would also be the case for explanations, as discussed by \cite{Berkovsky17,Kouki19,Hernandez20}. Since a main objective of providing explanations is to support users in their decision-making, investigating the effect of different personal styles to perform such a process is of particular interest to us. Particularly, we focus on the moderating effect of the \textit{rational} and \textit{intuitive} decision making styles \cite{Hamilton16}, the former characterized as a propensity to search for information and evaluate alternatives exhaustively, and the latter by a quick processing based mostly on hunches and feelings. Furthermore, since review-based explanations rely on the expressed opinions of other users, we investigated the effects of the extent to which users  are inclined to adopt the perspective of others when making decisions, a trait defined as \textit{social awareness} by \cite{Casel13}. We also considered \textit{visualization familiarity}, i.e. the extent to which a user is familiar with graphical or tabular representations of information. Consequently, \textbf{RQ3}: How do individual differences in decision-making styles, social awareness or visualization familiarity moderate the perception of review-based explanations with different degrees of interactivity and presentation styles?

Finally, the contributions of this paper can be summarized as follows: \begin{itemize}\item We formulate a scheme for explanations as interactive argumentation in review-based RS, inspired by dialogue models and formal argument structures. 
\item To test our research questions, we implemented an interface based on the proposed scheme, and a RS based on a matrix factorization model (i.e. EFM, \cite{Zhang14}), and sentiment-based aspect detection, using the state of art natural language processing model BERT (\cite{Devlin19}). 
\item We provide empirical evidence of the effect of review-based interactive explanations on users’ perception, as well as the influence of user characteristics on such perception.

\end{itemize}

\section{Related work}

Review-based explanatory methods leverage user generated content, rich in detailed evaluations on item features, which cannot be deduced from the general ratings, thus enabling the generation of more detailed explanations, compared to collaborative filtering (e.g. “Your neighbors’ ratings for this movie” \cite{Herlocker00}) and content-based approaches (e.g. \cite{Vig09}). Review-based methods allow to provide:
\textbf{1)} verbal summaries of reviews, using abstractive summarization from natural language generation (NLG) techniques  \cite{Carenini13,Costa18}, 
\textbf{2)} a selection of helpful reviews (or excerpts) that might be relevant to the user, detected using deep learning techniques and attention mechanisms \cite{Chen18,Donkers20}, 
\textbf{3)} a statistical view of the pros and cons of item features, usually using topic modelling or aspect-based sentiment analysis \cite{Wu15,Zhang14,Dong14}, information that is integrated to RS algorithms like matrix or tensor factorization \cite{Zhang14,Bauman17,Wang18b}) to generate both recommendations and aspect-based explanations. 

Our evaluation is based on the third approach, and is particularly related to the model proposed by \cite{Zhang14}, since it facilitates getting statistical information on users’ opinions, which has been proven to be useful for users \cite{Muhammad16,Hernandez20}, and can be provided in explanations with different presentation styles (strictly verbal or visual). Yet, the optimal way of presenting such information, either in a textual (short summaries) or a graphical form (e.g., different types of bar charts) remains unclear. In addition to information display factors, a second factor could also influence users' perception of the explanations: the possibility of interacting with the system, to better understand the rationale for its predictions. Interactive explanations have been already addressed in the field of explainable artificial intelligence (XAI) (although to a much lesser extent compared to static explanations \cite{Abdul18}). Here, the dominant trend has been to provide mechanisms to check the influence that specific features, points or data segments may have on the final predictions of a machine learning (ML) algorithm, as in the work of \cite{Krause16,Cheng19,Sokol20a}. However, despite the progress of XAI interactive approaches, their impact and possibilities in explainable RS remain largely unexplored, as well as the empirical validation of their effects on users. More specifically, the dominant ML interactive approach differs from ours in at least two ways: 1) we use non-discrete and non-categorical sources of information, subjective in nature and unstructured, which, however, can be used to generate both textual and visual structured arguments 2) such approach is designed to meet the needs of domain experts, i.e. users with prior knowledge of artificial intelligence, while we aim to target the general public. 

Therefore, we explore in this paper an interactive explanation approach that facilitates the exploration of arguments that support claims made by the system (why an item is recommended). To this end, we adopted the definition of interactivity stated by Steuer \cite{Steuer92}: “extent to which users can participate in modifying the form and content of mediated environment in real time”, and characterized the degree of interactivity of proposed explanations through the Liu and Shrum dimensions of interactivity \cite{Liu02}: active control and two-way communication. The first is characterized by voluntary actions that can influence the user experience, while the second refers to the ability of two parties to communicate to one another. Active control is reflected in our proposal by the possibility to use hyperlinks and buttons, that allow users to access explanatory information at different levels of detail at will, while two-way communication is represented by the ability to indicate the system (through pre-defined questions) which are their most relevant features, so the presentation of the explanatory content (answers) is adjusted accordingly.

In order to formulate and test an interactive flow for review-based explanations, we set our focus on argumentative models that may enable the two-way communication desideratum.  In contrast to static approaches to explanation, dialog models have been formulated conceptually \cite{Walton00,Arioua15,Madumal20,Rago20}, allowing arguments over initial claims in explanations, within the scope of an interactive exchange of statements. Despite the potential benefit of using these models to increase users' understanding of intelligent systems \cite{Miller18,Weld19}, their practical implementation in RS (and in XAI in general) still lacks sufficient empirical validation \cite{Sokol20,Miller18,Madumal20}. This dialogical approach contrasts with other argumentative - though static - explanation approaches \cite{Carenini06,Bader12,Lamche14,Zanker14,Hernandez20} based on static schemes of argumentation (e.g. \cite{Toulmin58,Habernal17}), where little can be done to indicate to the system that the explanation has not been fully understood or accepted, and that additional information is still required. Consequently, we formulated a scheme of interactive explanations in review-based RS, combining elements from dialogue models and static argument schemes (section 3), and conducted a user study to test the effect of the proposed interactive approach.

Effects of interactivity have been studied widely in fields like online shopping and advertising \cite{Liu02,Song08}, and more specifically in the evaluation of critique-based RS, where users are able to specify preferences for the system to recalculate recommendations, which has been found to be beneficial for user experience \cite{Chen12,Loepp14,Loepp15}. Despite the intuitive advantages that interactivity can bring, interactivity does not always translate into a more positive attitude towards the system, since it also depends on the context and the task performed \cite{Liu02}. Nevertheless, it has also been shown that higher active control is beneficial in environments involving information needs, and a clear goal in mind \cite{Liu02}, which is actually our case (i.e. deciding which hotel to book).
 
Furthermore, we hypothesized (in line with \cite{Liu02}) that a number of user characteristics may moderate the effect of interactive functionalities, on the perception of explanations. Particularly, we aimed to test the moderating effect of decision-making styles and social awareness. In regard to the former, research has shown that it is determined significantly by preferences and abilities to process available information \cite{Driver90}. Particularly, we believe that users with a predominant rational decision making style would better perceive explanations with a higher degree of interactivity, than explanations with less possibility of interaction, given their tendency to thoroughly explore information when making decisions \cite{Hamilton16}. On the other hand, more intuitive users may not find the interactive explanations very satisfactory, given their tendency to make decisions through a quicker process \cite{Hamilton16}, so that a first explanatory view would be sufficient, and it would not be necessary to navigate in depth the arguments that the system can offer. As for social awareness, and in line with results reported by \cite{Hernandez20}, we hypothesize that users with a higher social awareness may perceive explanations with higher interactivity more positively, given their tendency to take into account the opinions of others, and to adjust their own using those of others, while choosing between various alternatives \cite{Sniezek95}, which has been proved to be beneficial during decision making \cite{Yaniv07}, and is facilitated by our approach. 

Finally, in regard to presentation styles, visual arguments (a combination of visual and verbal information) may have a greater "rhetorical power potential" than verbal arguments, due (among others) to their greater immediacy  (possibility of quick processing) \cite{Blair12}. This could especially benefit users with a predominantly intuitive decision-making style, due to their usually quick manner of making decisions, based mostly on first impressions \cite{Hamilton16}. However, users with lower visual abilities might benefit less from a presentation based on images or graphics \cite{Schnotz14,Kirby88}. Consequently, we believe that when exposed to graphic-based explanation formats, higher interactive explanations may be beneficial to users with lower visual familiarity, as they could access additional information to better understand the explanations provided. 

\section{Scheme for explanations as interactive argumentation in review-based RS}

In order to evaluate our research questions, we designed an interaction scheme for the exploration of explanatory arguments in review-based RS. This scheme is inspired by dialog-based explanation models \cite{Walton04,Walton11,Madumal20}, in which instead of a single issue of explanatory utterances, an explanation process is regarded as an interaction, where a user could indicate when additional arguments are required, to increase their understanding of system claims. Walton \cite{Walton04,Walton11} modeled the possible actions as explanation requests or attempts, the former representing user questions, and the latter characterized as a set of assertions as system response. On the other hand, Madumal et al. \cite{Madumal20} noted that argumentation may occur within explanation, and modeled the shift between explanatory and argumentative dialogue, as well as the explanatory loops that can be triggered, when follow-up questions arise. While this type of models may help to define the moves allowed within an explanatory interaction, they offer little indication of how the arguments within the interaction moves should be structured, to increase their acceptance by users. To this end, we rely on the scheme by Habernal et al. \cite{Habernal17}, an adaptation of the Toulmin model of argumentation \cite{Toulmin58}, and formulated to better represent the kind of arguments usually found in user-generated content. This scheme involves: claim (conclusion of the argument), premise (a general reason to accept a claim), backing (specific information or additional evidence to support the claim), rebuttal (statement that attacks the claim) and refutation (statement that attacks the rebuttal). 

Our proposed scheme is shown in Figure 1. Unlike Walton, who modeled explanatory movements as explanation requests and attempts, we considered an explanation process as a sequence of \textit{argumentation attempts} (the system intends to provide arguments to explain something) followed by \textit{argument requests} (the user ask the system to provide - follow-up - arguments that support the claim that user will find the recommended item useful). The realization of such a scheme as a user interface developed for validation with users is depicted in Figure 2.

\begin{figure}[htp]
\centering
\includegraphics[width=140mm, height=50mm]{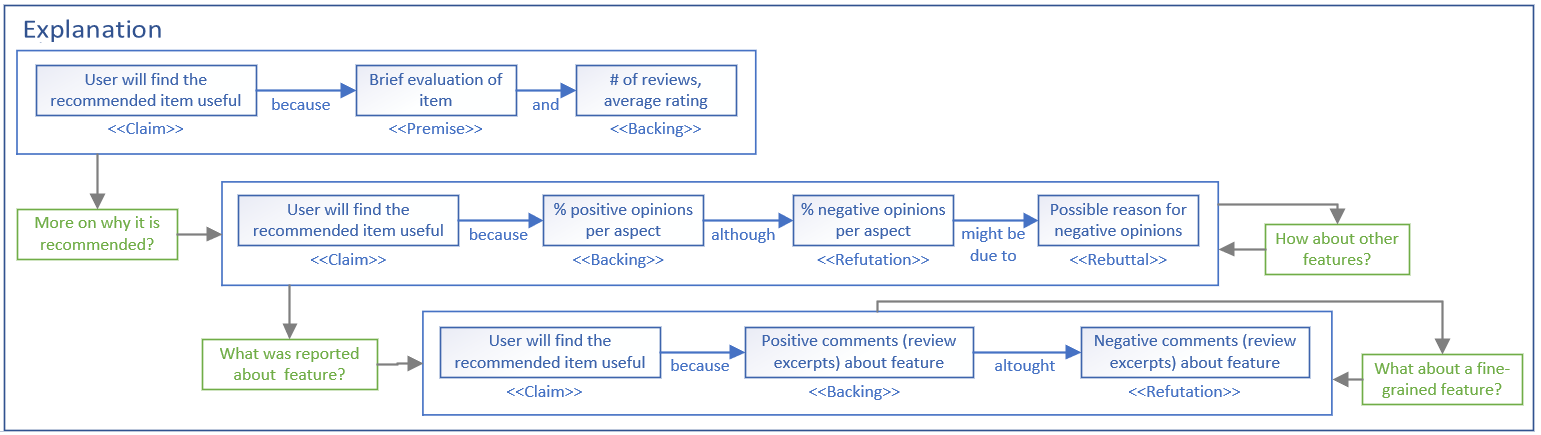}
\caption{Scheme for explanations as interactive argumentation in review-based RS. Blue boxes represent argumentation attempts by the system, green boxes the argument requests by users.}
\end{figure}

\begin{figure}[htp]
\centering
\includegraphics[width=65mm, height=87mm]{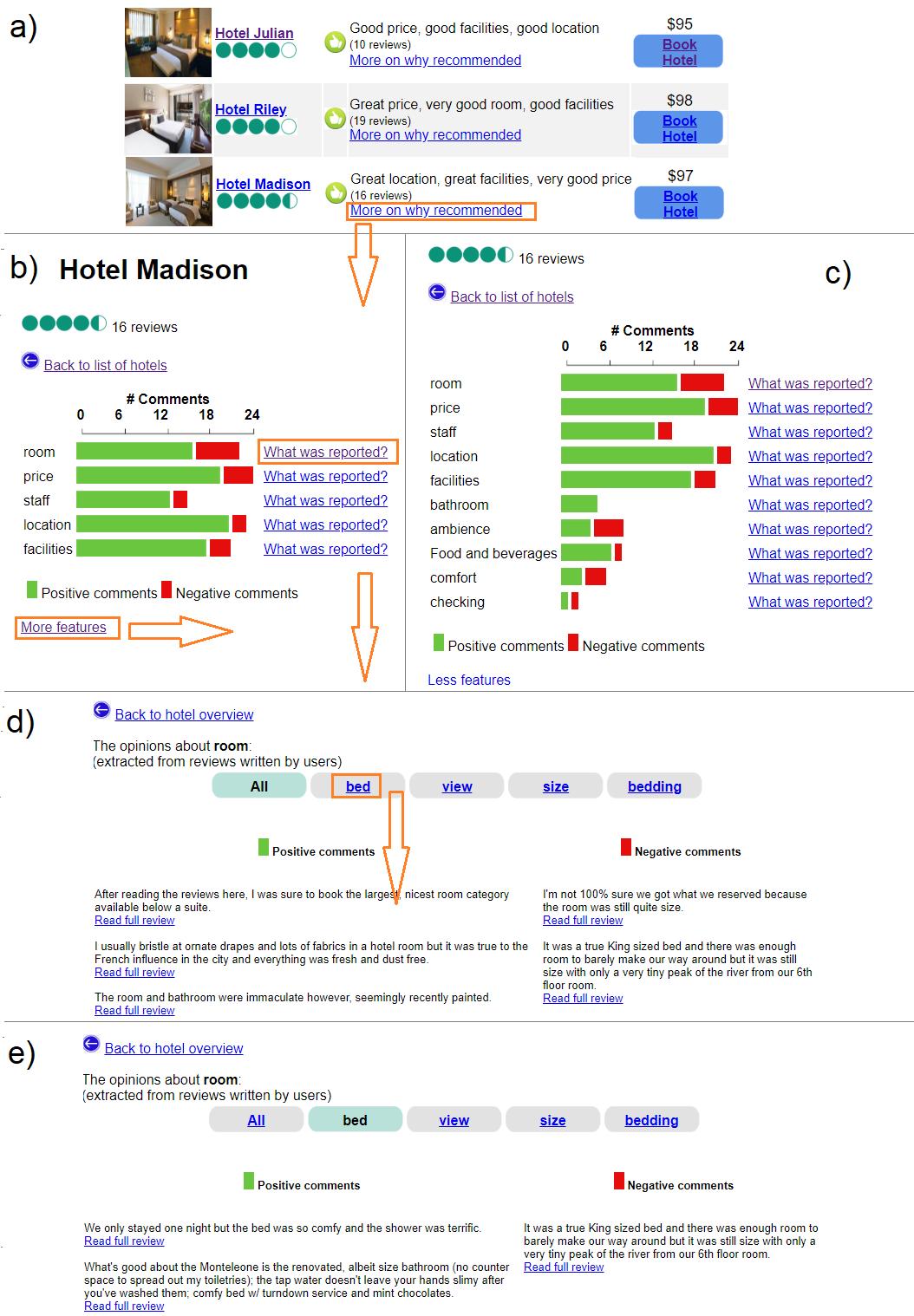}
\caption{Screenshots of implemented system for user study. Orange arrows depict the sequence of allowed moves, pointing towards the next interface provided. Steps c,d,e are enabled only in study condition interactivity “high”. a) List of recommended hotels, and first level argumentation attempt; when clicking on “More on why recommended”, system displays: b-c) a second level argumentation attempt; when clicking on “What was reported?”, system shows d) a third level argumentation attempt on the chosen feature; when clicking on an feature button, system shows e) only statements on the fine-grained chosen feature.}
\end{figure}

As noted by \cite{Miller18} and \cite{Madumal20}, an explanatory dialogue can take place both through verbal interactions and through a visual interface (non-verbal communication, or a combination of verbal and visual elements), which applies to both questions and answers. As for argument presentation styles, while arguments are usually associated with oral or written speech, arguments can also be communicated using visual representations (e.g. graphics or images) \cite{Blair12}. Thus, we considered the following styles for the argumentation attempt “\% of positive and negative opinions”: 1) Table (Figures 3a, 3b), bar chart (Figures 3c, 3d), and text (Figures 3e, 3f), the latter using the template proposed by \cite{Hernandez20}, which facilitates the display of rebuttal statements, which can hardly be represented graphically.
\squeezeup
\begin{figure}[htp]
\centering
\includegraphics[width=60mm, height=40mm]{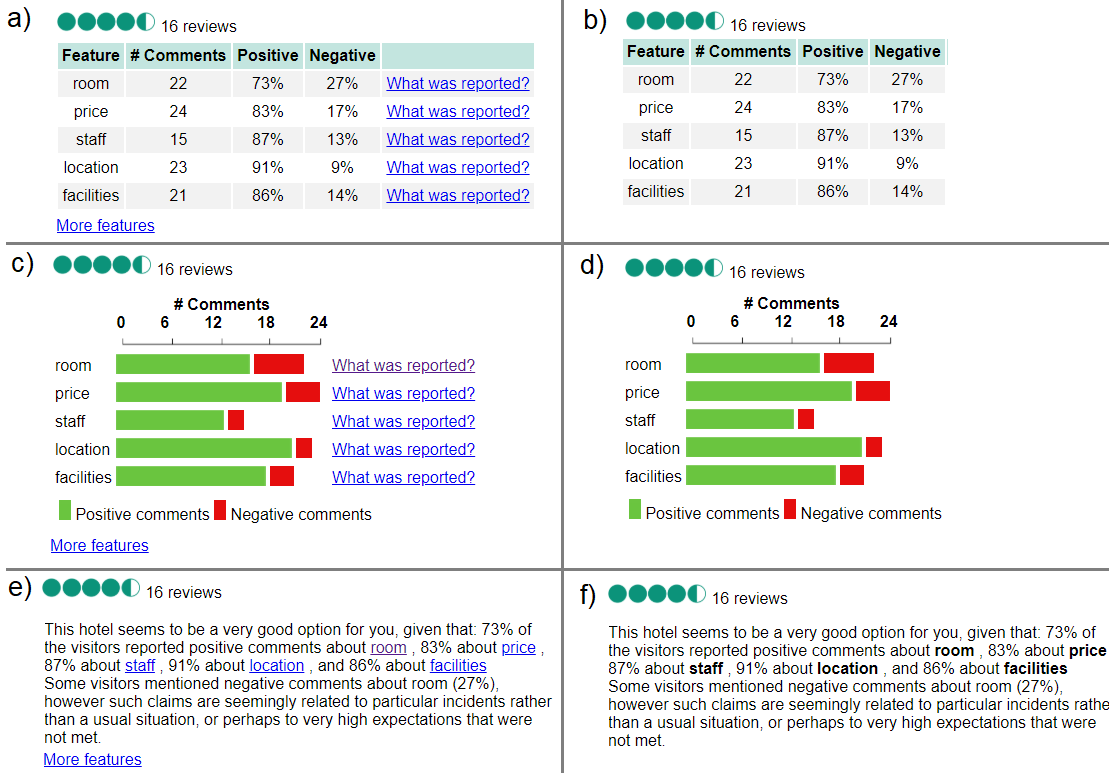}
\caption{Manipulation of \textit{presentation} style in combination with \textit{interactivity}, in user study. Top: \textit{table}, middle \textit{bar chart}, bottom \textit{text}. Left: interactivity \textit{high}, right: interactivity \textit{low}. }
\end{figure}
\squeezeup
\section{User study}
To answer our research questions, we implemented a RS that reflects the scheme described in section 3, and conducted a user study to compare users’ perception of the overall system (in terms of transparency, effectiveness and trust), and of the explanations (in terms of explanation confidence, transparency, persuasiveness, satisfaction and sufficiency). We considered the factor \textit{interactivity}, and the values “high” (users could make all possible argument requests (see Figures 1 and 2), and “low” (users could only make the initial argument request “more on why recommended?”). To validate the effect of factor \textit{presentation} style, we considered the values: \textit{table} (Figures 3 a,b), \textit{bar chart} (Figures 3 c,d) and \textit{text} (Figures 3 e,f). The study follows a 3x2 between-subjects design, and each participant was assigned randomly to one of six conditions (combination of {\itshape interactivity} and {\itshape presentation} style). We hypothesized:

\textbf{H1}: Users' perception of the system and its explanations is more positive when they are given explanations with higher interactivity.

\textbf{H2}: Users with a predominantly rational decision style perceive explanations with higher interactivity more positively than less rational decision makers.

\textbf{H3}: Less intuitive users perceive explanations with higher interactivity more positively, compared to more intuitive users.

\textbf{H4}: Users with greater social awareness perceive higher interactive explanations more positively than users with less social awareness. 

\textbf{H5a}: Users with a predominantly intuitive decision-making style or \textbf{H5b} a greater visualization familiarity will prefer bar chart explanations over text explanations, regardless of interactivity. 

\textbf{H6}: Users who are less familiar with data visualization will perceive explanations with higher interactivity more positively, particularly in the case of more challenging visualizations such as bar charts.

\subsection{Participants}

We recruited 170 participants (66 female, mean age 37.61 and range between 18 and 72) through Amazon Mechanical Turk. We restricted the execution of the task to workers located in the U.S, with a HIT (Human Intelligence Task) approval rate greater than 98\%, and a number of HITs approved greater than 500. We applied a quality check to select participants with quality survey responses (we asked validation questions to check attentiveness within questionnaires, and questions related to the content of the system). We discard participants with less than 10 (out of 12) correct answers, or no effective interaction with the system (checked in logs). The responses of 27 of the 197 initial participants were then discarded for a final sample of 170 subjects (statistical power of 90\%, $\alpha$ =0.05). Participants were rewarded with \$1.4 plus a bonus up to \$0.40 depending on the quality of their response to the question “Why did you choose this hotel?” set at the end of the survey. Time devoted to the task by participants (in minutes): M=10.88, SD= 1.62.

\subsection{Dataset and Implemented system}

\textit{Dataset and aspect annotation}: ArguAna \cite{Wachsmuth14}, includes hotel reviews and ratings from TripAdvisor; sentiment and explicit features are annotated sentence wise. We categorized the explicit features in 10 general features (room, price, staff, location, facilities, bathroom, ambience, food and beverages, comfort and checking), with the help of 2 annotators (Krippendorff's alpha of 0.72), aiming to train a classifier to detect the main aspect addressed in a sentence (e.g. “I loved the bedding” would be classified as \textit{room}).

\textit{Aspect-based sentiment detection}: We trained a BERT classifier \cite{Devlin19} to detect the general feature addressed within a sentence: we used a 12-layer model (\textit{BertForSequenceClassification}), 6274 training sentences, 1569 test sentences, F-score 0.84 (macro avg.). We also trained a BERT classifier to detect the sentiment polarity, using a 12-layer model (\textit{BertForSequenceClassification}), 22674 training sentences, 5632 test sentences, \textit{F}-score 0.94 (macro avg.). Classifier was used to \textbf{1)} consolidate the quality of hotels and relevance of aspects to users (see Figures 2b, 2d), and \textbf{2)} to present participants with negative and positive excerpts from reviews regarding a chosen feature (Figures 2d, 2e).

\textit{Explainable RS method}: We implemented the Explicit Factor Model (EFM) \cite{Zhang14}, a review-based matrix factorization (MF) method to generate both recommendations and explanations. The rating matrix consisted of 1284 items and 884 users extracted from the ArguAna dataset (only users with at least 5 written reviews were included), for a total of 5210 ratings. Item quality and user preferences matrices were consolidated using the sentiment detection described previously. The number of explicit features was set to 10. Model-specific hyperparameters were selected via grid-search-like optimization. After 100 iterations, we reached an RMSE of 1.27. Finally, values of predicted rating matrix were used to sort the list of recommendations and also shown within explanations (average hotel rating represented with 1-5 green circles). Values of the quality matrix were also used to calculate the percentages of positive and negative comments regarding different features (Figure 3). 

\textit{Personalization mechanism}: To reduce implications of the \textit{cold start} problem \cite{Schein02} (system does not have enough information about the user to generate an adequate profile and thus, personalized recommendations), participants were asked for the five hotel features that mattered most to them, in order of importance. The system calculated a similarity measure, to detect users within the EFM preference matrix with a similar order of preferences. Then the most similar user was used as a proxy to generate recommendations, i.e. we selected the predicted ratings of this proxy user, and used them to sort recommendations and features within explanations.

\subsection{Questionnaires}

\textit{Evaluation}: We utilized items from \cite{Pu11} to evaluate the perception of system transparency (construct \textit{transparency}, user understands why items were recommended), of system effectiveness \cite{Knijnenburg12} (internal reliability Cronbach’s $\alpha$ = 0.85, construct \textit{perceived system effectiveness}, system is useful and helps the user to make better choices), and of trust in the system \cite{McKnight02} ($\alpha$ = 0.90, constructs \textit{trusting beliefs}, user considers the system to be honest and trusts its recommendations; and \textit{trusting intentions}, user willing to share information). We used the user experience items (UXP) of \cite{Kouki19} to address explanations reception, which we will refer to as \textit{explanation quality} ($\alpha$ =  0.82), comprising: explanation confidence (user is confident that she/he would like the recommended item), explanation transparency (explanation makes the recommendation process clear), explanation satisfaction (user would enjoy a system if recommendations are presented this way), and explanation persuasiveness (explanations are convincing).  We added an item adapted from \cite{Donkers20} (explanations provided are sufficient to make a decision) to evaluate explanation sufficiency. All items were measured with a 1-5 Likert-scale (1: Strongly disagree, 5: Strongly agree). 

\textit{User characteristics}: We used all the items of the Rational and Intuitive Decision Styles Scale \cite{Hamilton16} (internal reliability Cronbach’s $\alpha$ = 0.84 and $\alpha$ = 0.92, respectively), the scale of the social awareness competency proposed by \cite{Casel13} ($\alpha$ = 0.70), and the visualization familiarity items proposed by \cite{Kouki19} ($\alpha$ = 0.86). All items were measured with a 1-5 Likert-scale (1: Strongly disagree, 5: Strongly agree). 

\subsection{Procedure}
Instructions indicated that a list of hotels reflecting the results of a hypothetical hotels’ search and within the same price range would be presented (i.e no filters to search hotels were offered to participants), and that they could click on the name of a desired hotel to see general information about it. However, we asked, as we were more interested in their views on the explanations given for each recommendation, to click on the "More on why recommended" links of hotels they might be interested in, and to explore the information provided. No further instructions were given regarding how to interact with the different interaction options, since we were interested to address to what extent the users used them or not. Users were instructed to indicate which hotel they would finally choose, and to write a few sentences reporting their reasons for it, for which a bonus up to \$0.4 would be paid, depending on the quality of this response, with the aim of achieving a more motivated choice by the participants, as well as to encourage a more effective interaction with the system. We then presented a cover story, which sought to establish a common starting point in terms of travel motivation (a holiday trip). Next, we presented to the participants the system showing a list of 30 recommended hotels (sorted by predicted rating), and their corresponding personalized explanations (system implementation details in section 4.2).
Finally, evaluation and validation questions were presented, plus an open-ended one, asking for general opinions and suggestions about the explanations. 

\subsection{Data analysis}

We evaluated the effect that interactivity and presentation style (independent variables IVs) may have on 2 different levels: \textbf{1)} \textit{overall system} perception, and \textbf{2)}  perception of specific aspects of  \textit{explanations}, and to what extent user characteristics (regarded as moderators or covariates) could influence such perception (rational and intuitive decision-making style, social awareness and visualization familiarity).

In case \textbf{1)} the dependent variables (DVs) are evaluation scores on: system transparency (user understands why items were recommended), effectiveness (system helps user to make better decisions), trust (user considers the system to be honest and trusts its recommendations) and explanation quality, a variable calculated as the average of scores reported on specific aspects of explanations: satisfaction, transparency, persuasiveness, confidence and sufficiency. 

In case \textbf{2)} the DVs are addressed explanation-wise: confidence (explanation makes user confident that she/he will like the recommendation), explanation transparency (explanation makes the recommendation process clear), satisfaction (user would enjoy a system if recommendations are presented this way), persuasiveness (explanations are convincing), and sufficiency (explanations provided are sufficient to make a decision). 

Scores of the rational and the intuitive decision making styles, social awareness and visualization familiarity for each individual as the average of the reported values for the items of every scale.  Internal consistency (Cronbach’s alpha) was checked for system evaluation and user characteristics constructs (see section 4.3).

\textit{Overall system perception}: Given that DVs are continuous and correlated (see Table 1), a MANCOVA analysis was performed. Subsequent ANCOVA were performed to test main effects of IVs and covariates, as well as the effect of interactions between them. Q-Q plots of residuals were checked to validate the adequacy of the analysis.

\textit{Perception of explanations}: DVs are ordinal (scores are the reported answers to single questionnaire items), thus we performed ordinal logistic regressions to test influence on DVs by predictor variables (IVs and covariates), no multicollinearity was tested, as well as Q-Q plots of residuals. DVs are also correlated (see Table 2), so significant tests were conducted using Bonferroni adjusted alpha levels of .01 (.05/5).

\textit{Use of interactive options}: Calculated based on system activity logs. A Mann-Whitney U test was used to compare distributions of users characteristics who used or not use such options.

\section{Results}

\subsection{Evaluation and User Characteristics Scores}

The average evaluation scores by presentation style and interactivity are shown in Tables 1 and 2. Distributions of the scores of rational (\textit{M} = 4.35, \textit{SD}= 0.50) and intuitive (\textit{M} = 2.59, \textit{SD}= 0.98) decision making styles, social awareness (\textit{M} = 4.04, \textit{SD}= 0.53) and visualization familiarity (\textit{M} = 3.23, \textit{SD}= 0.95) are depicted in Figure 4a.
\vspace*{-\baselineskip}
\begin{table*}[]
\caption{Mean values and standard deviations of perception on the overall system, per \textit{presentation} style and \textit{interactivity} (n=170), p$<$0.05*, p$<$0.01**; values reported with a 5-Likert scale; higher mean values correspond to a positive perception of the overall RS. Pearson correlation matrix, p$<$0.001 for all coefficients.}
\begin{tabular}{lrrrrlllrrrrllrll}
\cline{1-17}
\multicolumn{1}{r}{\textit{Presentation:}} & \multicolumn{2}{c}{Text} & \multicolumn{2}{c}{Table} & \multicolumn{2}{c}{Bar chart} & \begin{tabular}[c]{@{}l@{}}\textit{Inter-}\\\textit{activity:}\end{tabular} & \multicolumn{2}{c}{Low} & \multicolumn{2}{c}{High} & \textit{Corr:} & \multicolumn{4}{l}{Variable} \\ \cline{1-17}
\textit{Variable} & \multicolumn{1}{c}{M} & \multicolumn{1}{c}{SD} & \multicolumn{1}{c}{M} & \multicolumn{1}{c}{SD} & \multicolumn{1}{c}{M} & \multicolumn{1}{c}{SD} & & \multicolumn{1}{c}{M} & \multicolumn{1}{c}{SD} & \multicolumn{1}{c}{M} & \multicolumn{1}{c}{SD} & & 1 & \multicolumn{1}{l}{2} & 3 & 4 \\
1. Expl. Quality & 3.98 & 0.52 & 4.10 & 0.56 & 4.07 & 0.70 & & 3.92 & 0.61 & 4.17 & **0.55 & & & \multicolumn{1}{l}{} & & \\
2. Transparency & 4.14 & 0.52 & 4.11 & 0.86 & 3.91 & 0.99 & & 4.02 & 0.78 & 4.08 & 0.86 & & \multicolumn{1}{r}{0.51} & — & & \\
3. Effectiveness & 3.95 & 0.69 & 4.05 & 0.73 & 4.04 & 0.78 & & 3.91 & 0.79 & 4.11  & * 0.67 & & \multicolumn{1}{r}{0.75} & 0.56 & \multicolumn{1}{r}{—} & \\
4. Trust & 3.91 & 0.60 & 3.99 & 0.58 & 3.97 & 0.72 & & 3.86 & 0.67 & 4.05 & * 0.57 & & \multicolumn{1}{r}{0.74} & 0.55 & \multicolumn{1}{r}{0.79} & \multicolumn{1}{r}{—}
\end{tabular}
\end{table*}
\vspace*{-\baselineskip}
\begin{table*}[]
\caption{Mean values and standard deviations of perception on explanation specific aspects, per \textit{presentation} style and \textit{interactivity} (n=170), p$<$0.05*, p$<$0.01**; values reported with a 5-Likert scale; higher mean values correspond to a positive perception on the explanations. Pearson correlation matrix, p$<$0.001 for all coefficients.}
\begin{tabular}{lrrrrlllrrrrlllll}
\cline{1-17}
\multicolumn{1}{r}{\textit{Presentation:}} & \multicolumn{2}{c}{Text} & \multicolumn{2}{c}{Table} & \multicolumn{2}{c}{Bar chart} & \begin{tabular}[c]{@{}l@{}}\textit{Inter-}\\\textit{activity:}\end{tabular} & \multicolumn{2}{c}{Low} & \multicolumn{2}{c}{High} & \textit{Corr:} & \multicolumn{4}{l}{Variable} \\ \cline{1-17}
\textit{Variable} & \multicolumn{1}{c}{M} & \multicolumn{1}{c}{SD} & \multicolumn{1}{c}{M} & \multicolumn{1}{c}{SD} & \multicolumn{1}{c}{M} & \multicolumn{1}{c}{SD} & & \multicolumn{1}{c}{M} & \multicolumn{1}{c}{SD} & \multicolumn{1}{c}{M} & \multicolumn{1}{c}{SD} & & 1 & 2 & 3 & 4 \\
1. Expl. confidence & 4.09 & 0.55 & 4.11 & 0.65 & 4.05 & 0.85 & & 3.95 & 0.74 & 4.21 & * 0.62 & & & & & \\
2. Expl. transparency & 4.16 & 0.73 & 4.19 & 0.83 & 4.16 & 0.86 & & 4.05 & 0.84 & 4.29 & * 0.77 & & \multicolumn{1}{r}{0.60} & \multicolumn{1}{r}{—} & & \\
3. Expl. satisfaction & 3.84 & 0.85 & 4.09 & 0.79 & 4.11 & 0.80 & & 3.88 & 0.84 & 4.14 & * 0.77 & & \multicolumn{1}{r}{0.40} & \multicolumn{1}{r}{0.53} & \multicolumn{1}{r}{—} & \\
4. Expl. persuasiveness & 3.84 & 0.71 & 3.96 & 0.71 & 3.93 & 0.82 & & 3.82 & 0.71 & 4.00 & 0.77 & & \multicolumn{1}{r}{0.64} & \multicolumn{1}{r}{0.47} & \multicolumn{1}{r}{0.45} & \multicolumn{1}{r}{—} \\
5. Expl. sufficiency & \multicolumn{1}{l}{3.96} & \multicolumn{1}{l}{0.79} & \multicolumn{1}{l}{4.14} & \multicolumn{1}{l}{0.81} & 4.09 & 0.83 & & \multicolumn{1}{l}{3.89} & \multicolumn{1}{l}{0.87} & \multicolumn{1}{l}{4.23} & \multicolumn{1}{l}{**0.71} & & 0.40 & 0.44 & 0.50 & 0.45
\end{tabular}
\end{table*}
\vspace*{-\baselineskip}
\subsection{Overall System Perception}

\textit{Interactivity}:
We found a significant multivariate effect of interactivity on overall system perception \textit{F}(4,157) = 2.68, \textit{p} = .034. Univariate tests revealed that interactivity significantly influences the perception of explanation quality \textit{F}(1,168) = 9.76, \textit{p} = .002, effectiveness \textit{F}(1,168) = 4.02, \textit{p} = .047, and trust \textit{F}(1,168) = 4.63, \textit{p} = 0.033. In all these cases, the average of every variable was significantly higher for the \textit{high} condition than for \textit{low} condition (see Table 1). 

\textit{Presentation style}:
We found no significant main effect of \textit{presentation} style.

\textit{Rational decision-making style}:
We found a significant multivariate effect of rational style, \textit{F}(4,157) = 7.55, \textit{p} $<$ .001. Univariate tests revealed a main effect of rational decision-making style on explanation quality, \textit{F}(1,168) = 20.27, \textit{p} $<$ .001, system transparency \textit{F}(1,168) = 8.25, \textit{p} = .005, effectiveness, \textit{F}(1,168) = 26.76, \textit{p} $<$ .001 and trust, \textit{F}(1,168) = 24.94, \textit{p} $<$ .001. In all these cases, a positive trend was observed between these variables and the rational decision-making style, i.e. the higher the rational decision-making score, the higher the perceived explanation quality, the transparency, the effectiveness and the trust, independent of the style or interactivity (see Figure 4b).

\textit{Social awareness}:
We found a significant multivariate effect of social awareness, \textit{F}(4,157) = 6.41, \textit{p} $<$ .001. Univariate tests revealed a main effect of social awareness on explanation quality \textit{F}(1,168) = 17.25, \textit{p} $<$ .001, system transparency \textit{F}(1,168) = 12.57, \textit{p} $<$ .001, effectiveness \textit{F}(1,168) = 22.85, \textit{p} $<$ .001 and trust \textit{F}(1,168) = 18.02, \textit{p} $<$ .001. In all these cases, a positive trend was observed between these variables and social awareness, i.e. the higher the social awareness score, the higher the perceived explanation quality, the transparency, the effectiveness and the trust, independent of the style or interactivity (see Figure 4c). 
\squeezeup
\begin{figure*}[htp]
\centering
\includegraphics[width=\textwidth,scale=0.5]{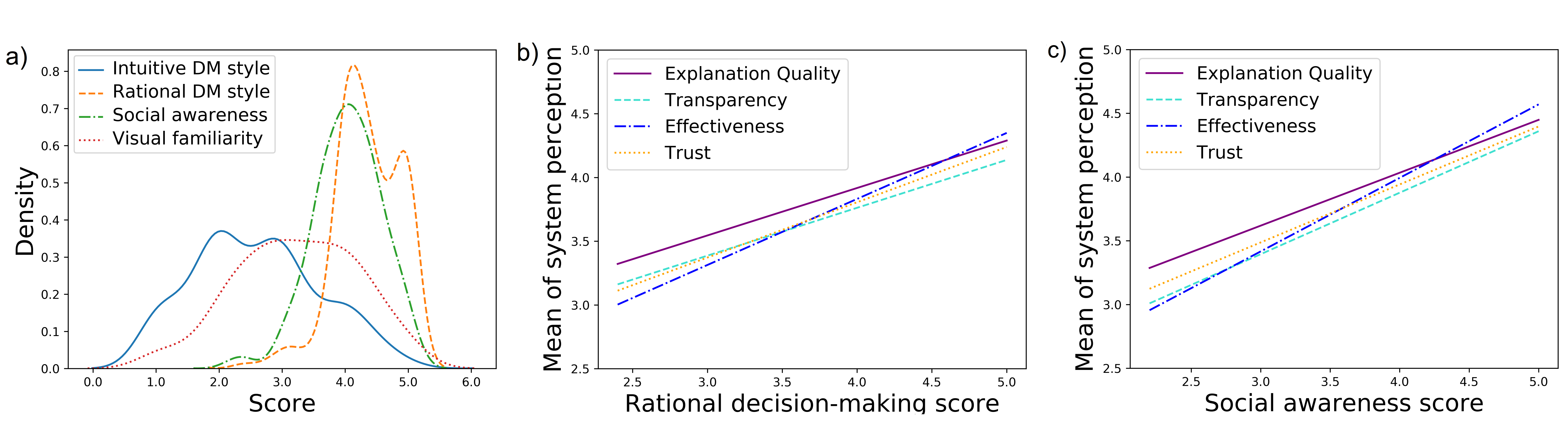}
\caption{a) Kernel density estimate of user characteristics scores: rational and intuitive decision making styles, social awareness and visualization familiarity. b) Effect of rational decision-making style on the perception of the overall system (fitted means of individual scores). c) Effect of social awareness on the perception of the overall system (fitted means of individual scores).}
\end{figure*}
\squeezeup
\subsection{Perception of Explanations}

\textit{Interactivity}:
We found a main significant  effect of interactivity; here, the odds of participants reporting higher values of explanation sufficiency when interactivity \textit{high} was 2.30 (95\% CI, 1.26 to 4.29) times that of interactivity \textit{low}, a statistically significant effect, Wald {$\chi$}2(1) = 7.32, \textit{p} = .007.
We observed a similar pattern in relation to explanation confidence (\textit{p} = .017), explanation transparency (\textit{p} = .043) and explanation satisfaction (\textit{p} = .041). However, this association (despite \textit{p} $<$ .05) is non-significant after Bonferroni correction (corrected \textit{p} $<$ 0.01).

\textit{Presentation style}:
We found no significant main effect of \textit{presentation} style.

Additionally, we observed a possible interaction (\textit{p}$<$= 0.05, although non-significant after Bonferroni correction, corrected \textit{p}$<$ 0.01) between:

\textit{Rational decision-making style and interactivity}:
An increase in rational decision-making score was associated with an increase in the odds of participants under interactive \textit{high} condition reporting higher values of explanation sufficiency, with an odds ratio of 3.20 (95\% CI, 0.99 to 10.65), Wald {$\chi$}2(1) = 3.81, \textit{p} = .051 (Figure 5a).

\textit{Intuitive decision-making style and presentation style}:
An increase in intuitive decision-making score was associated with an increase in the odds of participants under \textit{bar chart} condition reporting higher values of explanation satisfaction, with an odds ratio of 2.40 (95\% CI, 1.14 to 5.18), Wald {$\chi$}2(2) = 5.67, \textit{p} = .023, compared to participants under \textit{text} condition (see Figure 5b).

\textit{Social awareness and interactivity}:
An increase in social awareness score was associated with an increase in the odds of participants under interactive \textit{high} condition reporting higher values of explanation persuasiveness, with an odds ratio of 3.83 (95\% CI, 1.20 to 12.34), Wald {$\chi$}2(1) = 5.17, \textit{p} = .023 (Figure 5c).

\textit{Visualization familiarity and interactivity}:
An increase in visualization familiarity score was associated with an increase in the odds of participants under interactive \textit{high} condition reporting higher values of explanation satisfaction, odds ratio of 1.91 (95\% CI, 1.03 to 3.58), Wald {$\chi$}2(1) = 4.24, \textit{p} = .039 (Fig. 5d).

\begin{figure*}[h]
\centering
\includegraphics[width=49mm,scale=0.5]{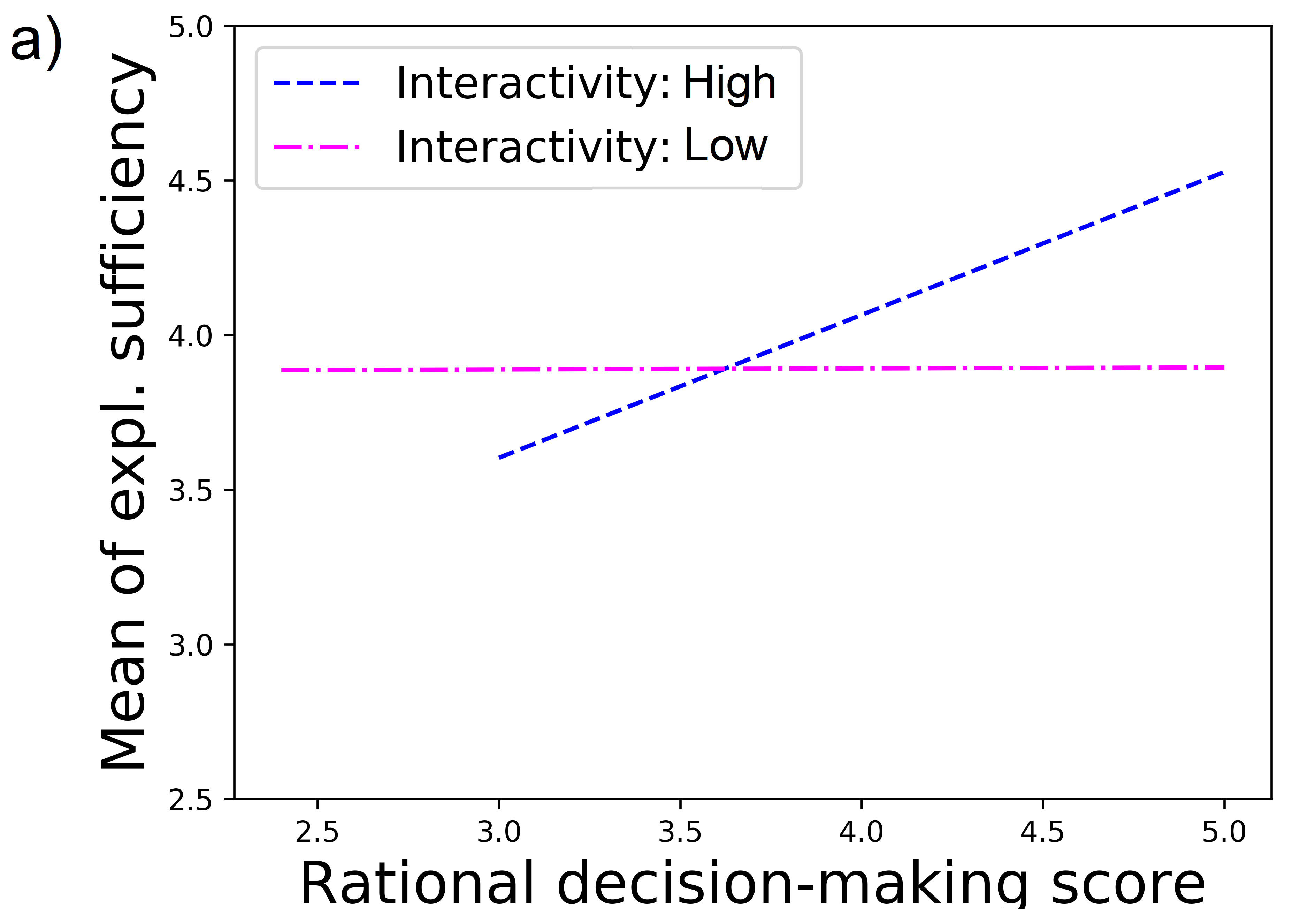}
\includegraphics[width=49mm, scale=0.5]{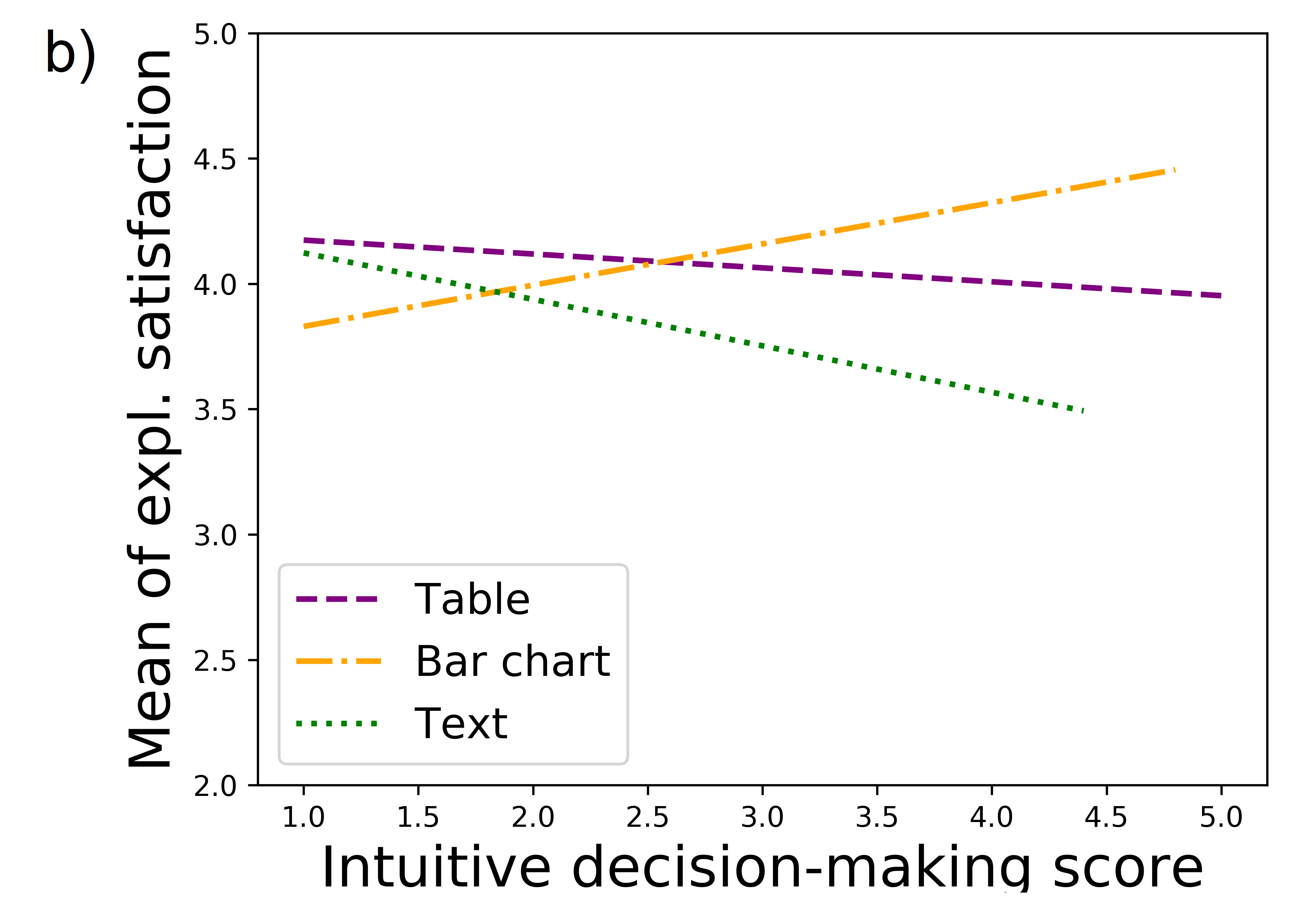}
\includegraphics[width=49mm,scale=0.5]{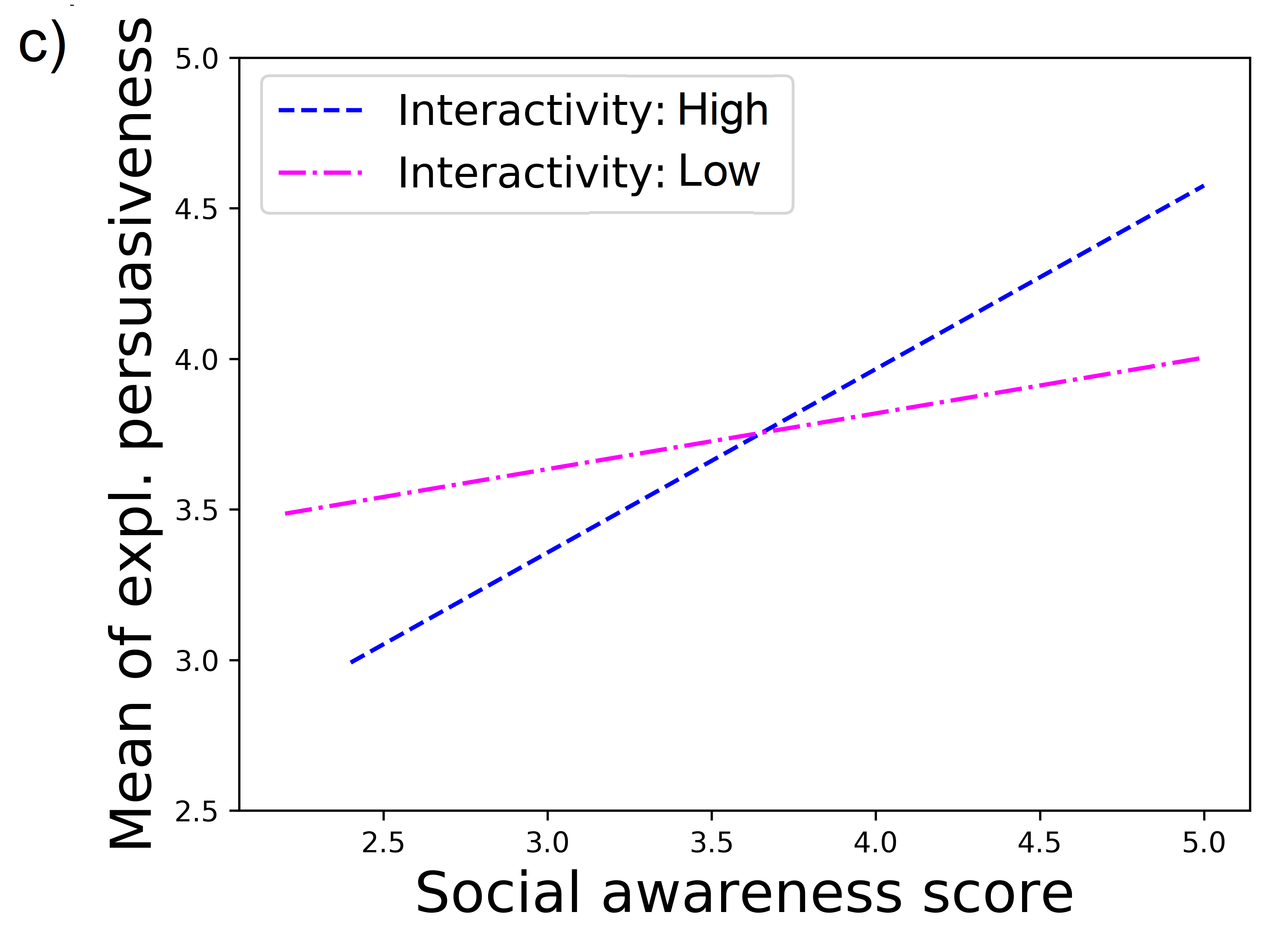}
\includegraphics[width=49mm, scale=0.5]{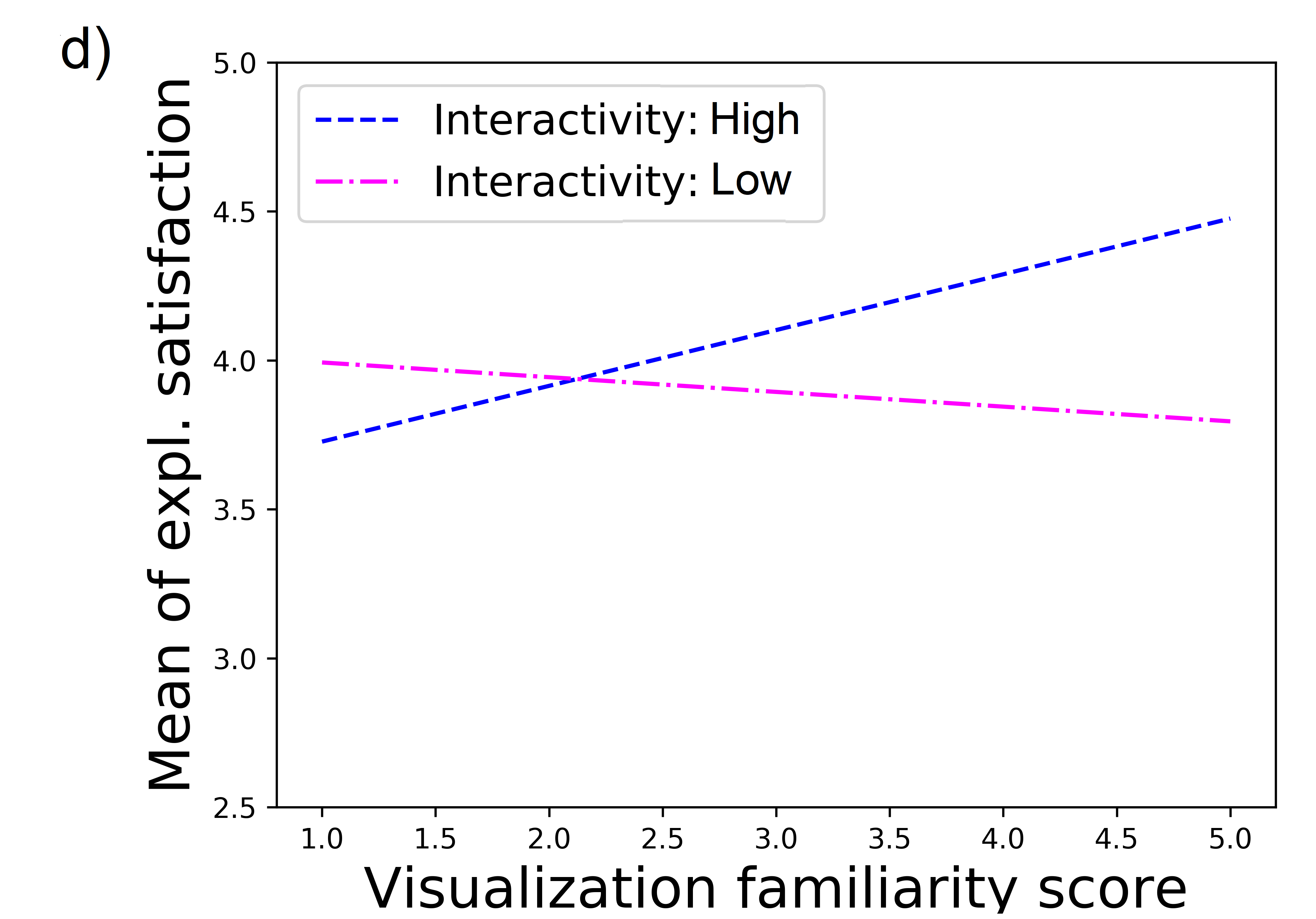}
\caption{Interaction plots (fitted means of individual scores) for perception of explanation:  a) sufficiency, interaction between interactivity and rational decision-making style. b) satisfaction, interaction between presentation and intuitive decision-making. c) persuasiveness, interaction between interactivity and social awareness. d) satisfaction, interaction between interactivity and visualization familiarity. }
\end{figure*}

\subsection{Use of interaction options}

48\% of the users assigned to the interactivity \textit{high} conditions used at least one of the interaction options provided. 48.15\% of participants used the ‘more features’ option when explanations were displayed using table, 26.92\% using bar chart and 33.3\% using text. 55.56\% of participants used the ‘what was reported’ option when explanations were displayed as table, 50\% as bar chart and 3.7\% as text. And 22.22\% of participants used the ‘comments on specific features’ option when explanations were displayed as table, 19.23\% as bar chart and 3.7\% as text.  Additionally, a Mann-Whitney U test revealed that the average of visualization familiarity scores of users who used the interaction options (\textit{M} = 2.98, \textit{SD} = 1.05) is significantly lower than the score of those that did not use them (\textit{M} = 3.41, \textit{SD} = 0.85) , \textit{U}(N\textsubscript{used}=41, N\textsubscript{not used}=44) = 678.50, \textit{p} = .024).

\section{Discussion}
Our results show that greater interactivity has a significantly positive effect on users’ perception, in terms of system effectiveness and trust, as well as of explanation quality, compared to explanations with lower interactivity, thus confirming our \textbf{H1}. We believe that the interactivity aspects addressed in our proposal could play a determining role in the observed effect, namely: active control and two-way communication. The former, by enabling users to be in control of which argumentative content to display; the latter by enabling them to indicate the system which argumentative statements require further elaboration, and which features are of real relevance at the time of making the decision, an approach that might contribute to a better acceptance and understanding of explanations, as predicted by dialog models of explanation \cite{Walton11,Hilton90}. 

However, the benefit and actual use of interactive options in review-based explanations might be influenced by individual differences, as discussed by \cite{Liu02} for the scope of online advertising and shopping. In particular, we found that the way people process information when making decisions would play an important role in the perception of interactive review-based explanations. More precisely, and in line with \textbf{H2}, we found that greater interactivity might have a more positive effect on the perception of explanation sufficiency by more rational users, which is explained by the propensity of people with a predominant rational decision-making style, to search for information and evaluate alternatives exhaustively \cite{Hamilton16}. However, and contrary to our expectations, we observed that the degree of intuitive decision style did not moderate the effect of interactivity on users' perception, so we cannot confirm our \textbf{H3}. In this regard, despite the predominant quick process based mostly on hunches that characterize more intuitive decision-makers \cite{Hamilton16}, we believe that looking at some verbatim excerpts from other users' reviews may also be of benefit to them, as they could corroborate whether their hunches are aligned with the system's assertions, although they may not do so as extensively as less intuitive users would do.

Additionally, in line with our \textbf{H4} and results reported by \cite{Hernandez20}, we observed that social awareness might moderate the effect of interactivity on explanation persuasiveness. Here, results suggest that participants with a higher disposition to listen and take into account others’ opinions, tend to perceive higher interactive explanations as more persuasive, which seems a consequence of the possibility to read reports of personal experiences by customers, who have already made use of the recommended items. This represents a potential advantage in the evaluation of experience goods like hotels, which is characterized by a greater reliance on word-of-mouth \cite{Nelson81,Klein98}. 

In line with \textbf{H5a}, our observations suggest that intuitive decision style might mediate the effect of presentation on explanation satisfaction, independent of interactivity. Particularly, explanatory arguments presented as a bar chart seemed to be perceived as more satisfactory to more intuitive users, than the presentation using a table or only text, presumably due to their greater immediacy \cite{Blair12}, thus facilitating the rapid decision-making process that characterizes more intuitive users. However, contrary to our expectations, we cannot conclude that users with more visualization familiarity will perceive the bar chart explanations better than the text-based ones (\textbf{H5b}). One possible reason could be that a text-based format makes it easier to visualize argumentative components as rebuttal and refutation, which could lead to a higher acceptance of an argument, as advocated by argumentation theory (\cite{Habernal17}), but could hardly be expressed through graph-based formats.

Additionally, although users with lower visualization familiarity tended to use the interaction options more,  we cannot confirm our hypothesis that those users would perceive graphic-based explanations (i.e. bar chart) better when more interactive options are offered, (\textbf{H6}). Actually, we found that users with more experience with data visualization reported a more positive perception for explanations with higher interactivity, independent of presentation style. We believe this is not due to difficulties understanding the explanations (as we thought would be the case for users with less visualization familiarity), but because higher interactivity facilitated a structured navigation and more appealing display of the data, which would not be as easy to process or useful if presented on a single static explanation. 

Overall, we observed a main effect of rational decision-making style and social awareness in the perception of the system and all the proposed explanations. This suggests that review-based explanations seem to benefit more the users who tend to evaluate information thoroughly and take into account the opinions of others when making decisions, compared to users who use a more shallow information-seeking process.

\textit{Interactivity and transparency perception}. Despite the main effect of interactivity on the overall perception of the system and its explanations, the mean perception of system transparency (user understands why items were recommended) is only slightly higher for the interactivity \textit{high} condition than for the \textit{low} condition. We believe that the reason might be two-fold: 1) Walton's \cite{Walton11} suggests to include an explicit mechanism to confirm effective understanding by the user, so that if this has not yet been achieved, the iterative cycle of user questions and system responses may continue. In consequence, we believe that a more flexible approach in which the user could, for example, write their own questions, rather than the bounded link-based options, might contribute in this regard. And 2) users may be also interested in understanding the reasons why the hotel x is better than hotel y. This would not only be in line with the view of authors who claim that the \textit{why-questions} ask for a contrastive explanation (“why P rather than Q?”) \cite{Hilton90,Lipton90,Miller18}, but also concurs with some participants’ suggestions, that options for comparison would be very useful, e.g. “It'd be easier if information wasn't on each separate page, too. I'd like an option to compare and contrast options”.

\textit{Use of interaction options}. We observed that almost half of participants under the condition interactivity “high” actually used the interaction options, although participants were  not explicitly instructed to use them, so it can reasonably be inferred that their use was mainly voluntary. It is critical, however, that these options are named appropriately, indicating clearly their destinations (as stated by \cite{Farkas00} guidelines), to increase the probability of their use, as evidenced by the lack of use of the option to read reviews excerpts in the \textit{text} condition (Fig. 3e).

Additionally, some of the users assigned to the \textit{low} interactivity condition pointed to 1) the lack of access to additional information in connection to the explanations (particularly customer reviews) as a disadvantage, with about a quarter of those participants writing suggestions on the subject,  e.g. “I would prefer to read the actual reviews and understand why ratings were what they were”, or 2) insufficiency of aggregated percentages of positive and negative opinions to adequately explain recommendations, e.g. “I feel they maybe could have a lot more information more on SPECIFICALLY what they say about the room instead of just an overall aggregation”. In this regard, it is important to note though, that participants of all conditions had access to the full hotel reviews (they were included in the general view of each hotel).

\textit{Practical implications}. 
Our approach was specifically tested in hotels domain, however, since it allows users to navigate from aggregated accounts of other users’ opinions to detailed extracts of individual reviews, we believe it might generalize adequately to domains that involve the evaluation of experience goods, i.e. those whose attributes can only be evaluated when the product has already been experienced \cite{Nelson70}, and where the search for information is characterized by a greater reliance on word-of-mouth \cite{Nelson81,Klein98} for example restaurants, movies or books. Additionally, our findings lead to the following practical implications, to be considered in the design of review-based explanations in RS involving experience goods: 
\begin{itemize} 
\item Providing interactive explanations resembling an argumentative communication between system and user could contribute to a better perception of the system. This could be implemented using web navigation options, e.g. links or buttons that indicate explicitly their destination, indicating if possible, a \textit{why} or \textit{what} question that will be answered by the system afterwards.
\item Presenting both aggregated opinion statistics and excerpts of comments filtered by feature, as part of an interactive explanation, is a beneficial way to provide explanations sufficient in content, while avoiding overwhelming users with irrelevant data in a single step or screen.
\item Given the practical difficulty of detecting user characteristics (e.g., decision-making style or visualization familiarity) by the system, we suggest interactive options to be considered, not only to provide in-depth arguments or to detect the relevance of features to the user, but also to modify the presentation style of argument components.
\end{itemize}

\section{Conclusions and future work}

In this paper, we have presented a scheme for explanations as interactive argumentation in review-based RS, inspired by dialogue explanation models and formal argument schemes, that allows users to navigate from aggregated accounts of other users’ opinions to detailed extracts of individual reviews, in order to facilitate a better understanding of the claims made by the RS. We tested an implementation of the proposed scheme in the hotels domain, and found that more interactive explanations contributed to a more positive perception of effectiveness and trust in the system. We also found that individual differences in terms of user characteristics (e.g. decision-making style, social awareness and visualization familiarity) may lead to differences in the perception of the proposed implementation.

While our proposal suggests a first step towards an effective implementation of interactive explanations for review-based RS, some important improvements can still be considered, to increase users' perception of transparency, as pointed out in the previous section. Here, the provision of links with predefined \textit{why}, \textit{how} or \textit{what} questions, while practical, could be improved, for example, with the possibility for the user to ask more flexible questions, even in natural language. Thus, as future work, we plan to leverage advances of conversational agents (i.e. chatbots), natural language processing and natural language generation techniques, such as question answering and automatic summarization, to enhance the implementation proposed in this paper.

It is important to note that our scheme entails an explanatory dialogue on a single-item level. However, we plan in the future to investigate the effect of contrastive dialogue-based explanations of the type “Why P rather than not-P?”. In this respect, we believe that this type of explanation can be leveraged to enable users further possibilities to influence the recommendation process itself, e.g. requesting for a more refined set of recommendations that better suit their preferences, based on an explanatory contrast between the different options. The above might result in greater satisfaction with the overall system, as has been proven with interactive RS in the past, but this time from the explanations as such.

\subsubsection*{Acknowledgements}This work was funded by the German Research Foundation (DFG) under grant No. GRK 2167, Research Training Group “User-Centred Social Media”.

\bibliography{explanations-base}  
\bibliographystyle{splncs04} 

\end{document}